\documentclass[twocolumn]{aa}  

\usepackage{amsmath}
\usepackage{natbib}
\usepackage{graphicx}
\usepackage{txfonts}
\usepackage[usenames,dvipsnames,svgnames,table]{xcolor}
\usepackage[breaklinks=true]{hyperref}
\usepackage[all]{hypcap}
\hypersetup{
    colorlinks,
    linkcolor={red!50!black},
    citecolor={blue!50!black},
    urlcolor={blue!80!black}
}

\DeclareMathOperator{\sinc}{sinc}

\begin{document}

   \title{Tied-array holography with LOFAR}

   \author{P.~Salas\inst{1}\fnmsep\thanks{psalas@nrao.edu}\thanks{Current address: Green Bank Observatory, Green Bank, WV 24944, USA}
          \and
          M.~A.~Brentjens\inst{2}
          \and
          D.~D.~Bordenave\inst{3}
          \and
          J.~B.~R.~Oonk\inst{1,2}
          \and
          H.~J.~A.~R\"ottgering\inst{1}
          }

   \institute{Leiden Observatory, Leiden University, P.O. Box 9513, NL-2300 RA Leiden, The Netherlands             
              \and
              ASTRON, Netherlands Institute for Radio Astronomy, Oude Hoogeveensedijk 4, 7991 PD, Dwingeloo, The Netherlands
              \and
              Department of Astronomy, University of Virginia, P.O. Box 400325, 530 McCormick Road, Charlottesville, VA 22904-4325, USA
             }

%    \date{Received March XX, 2018; accepted April XX, 2018}
\date{\today}

  \abstract
   {A radio interferometer uses time delays to maximize its response to radiation coming from a particular direction.
   These time delays compensate for differences in the time of arrival of the wavefront at the different elements of the interferometer, and for delays in the instrument's signal chain.
   If the radio interferometer is operated as a phased array (tied array), the time delays cannot be accounted for after an observation, so they must be determined in advance.}
   {Our aim is to characterize the time delays between the stations in the core of the LOw Frequency ARray (LOFAR).}
   {We used radio holography to determine the time delays for the core stations of LOFAR (innermost $3.5$~km).
   Using the multibeaming capability of LOFAR we map the voltage beam faster than with a raster scan, while simultaneously calibrating the observed beam continuously.}
   {For short radio holographic observations ($60$~s and $600$~s) of 3C196, 3C147, and 3C48 we are able to derive time delays with errors of less than one nanosecond.
   After applying the derived time delays to the beamformer, the beam shows residuals of less than $20\%$ with respect to the theoretical beam shape.}
   {Tied-array holography could be a way towards semi-real-time beam calibration for the Square Kilometer Array.}


  \maketitle
%
%-------------------------------------------------------------------

\section{Introduction}

A radio telescope works by combining the signals received by the elements that constitute its aperture (a reflecting surface in the case of a dish, or an array of antennas in the case of a phased array).
In order to maximize the sensitivity of the telescope towards a particular direction, the signals arriving from that direction must be combined in phase, i.e., the time difference between the signals received by different aperture elements must be zero.
In the case of a dish this is accomplished by shaping the reflecting surface in such a way that all the signals arrive at the receiver at the same time; in a phased array it is done by introducing instrumental time delays between its elements to compensate for the time of arrival of the signal at the antennas \citep[e.g.,][]{Thompson2017}.

Deviations from a perfect phase alignment when the signals are combined lead to a loss in the efficiency of the telescope \citep[e.g.,][]{Ruze1952,Ruze1966,D'Addario2008}.
These phase misalignments can be caused by the telescope itself, or they can be produced in the path between the source of the signals and the telescope.
An example of the former are phase differences caused by misaligned panels in a reflector \citep[e.g.,][]{Baars2007} or by uncorrected cable delays in a phased array.

Different methods to reduce phase misalignments between the elements of an aperture have been developed.
These include photogrammetric measurements \citep[e.g.,][]{Wiktowy2003}, direct measurement of the aperture distribution \citep[e.g.,][]{Chen1998,Naruse2009}, holographic measurements \citep[e.g.,][]{Napier1973,Bennett1976,Scott1977,Baars2007,Hunter2011}, and calibration using astronomical sources \citep[e.g.,][]{Fomalont1999,Intema2009,Thompson2017,Rioja2018}.
This paper focuses on the holographic measurement of the aperture illumination of a large phased array telescope.

Since the work of \citet{Scott1977}, holographic measurements have been used to calibrate the dishes of the Very Large Array \citep[VLA, e.g.,][]{Kesteven1993,Broilo1993}, the Atacama Large Millimeter Array \citep[ALMA, e.g.,][]{Baars2007}, and the Green Bank Telescope \citep[GBT, e.g.,][]{Hunter2011}; to study the primary beam response of the Westerbork radio telescope \citep[WSRT, e.g.,][]{Popping2008} and the Allen telescope array dishes \citep[ATA, e.g.,][]{Harp2011}; and to characterize the beam and aperture of the LOw Frequency ARray (LOFAR) stations \citep[Brentjens et al. in prep.][]{}.
All these measurements have been restricted to the study of apertures $\lesssim100$~m in diameter.

In the regime of low frequencies and large apertures, holographic measurements are particularly challenging.
At low frequencies the ionosphere will introduce additional time delays depending on its total electron content \citep[TEC, e.g.,][]{Intema2009}.
To accurately measure the intrinsic phase errors between the elements of the phased array without ionospheric distortion, the phased array must be smaller than the diffractive scale of the ionosphere.
Night time observations of the ionosphere at $150$~MHz show that its diffractive scale is between $30$ and $3$~km \citep{Mevius2016}.

LOFAR operates at frequencies between $10$~MHz and $240$~MHz \citep[][]{vanHaarlem2013}.
This frequency range is covered by two different types of antennas: low band antennas (LBA, $10$--$90$~MHz) and high band antennas (HBA, $120$--$240$~MHz).
The HBA antennas are combined in a $4\times4$ tile with an analog beamformer.
The antennas and tiles are grouped into stations, and the stations are further combined to form an array.
For the core stations of LOFAR, the LBA stations consist of $96$ antennas, while the HBA stations have $48$ tiles split into two fields.
Of the $96$ antennas in a core LBA station the available electronics permits only $48$ to be actively beamformed.
There are $24$ stations in the core of LOFAR.
The core stations are connected via fiber to a central clock, thus their signals can be added coherently to form a telescope with a maximum baseline of $3.5$~km.
The stations in the innermost $350$~m are known as the \textit{Superterp}.

Each LOFAR LBA dipole observes the entire sky, while the HBA tiles have a field of view (FoV) of $30\degr$ at $150$~MHz.
Since the signals from the antennas and tiles are combined digitally, the stations can simultaneously point in multiple directions within their FoV \citep[e.g.,][]{Barton1980,Steyskal1987}.
When the signals from different stations are added together coherently, a phased array (known as a tied array) is formed. 
This enables LOFAR to form multiple tied-array beams (TABs) that point in different directions.

\section{Method}
\label{sec:method}

We want to determine the time delays for the array formed by the stations in LOFAR's core.
We refer to the tied array formed by these stations as the array under test (AUT).
In order to determine the time delays, we start from a map of its complex-valued beam $B$.
The basic procedure used to measure $B$ is the same as that employed by \citet{Scott1977}, with a difference in its implementation.
In their work, a raster scan was used to map the region around the bright unresolved source.
Here, we take advantage of LOFAR's multi-beaming capability to map the region around the bright unresolved source.
Using multiple TABs the whole region is mapped simultaneously, and there is always a TAB pointing towards the bright unresolved source.
In addition to speeding up the process by a factor equal to the number of simultaneous beams, this allows continuous calibration of the AUT and the reference stations by always having a TAB at the central calibrator source.

At the frequencies at which LOFAR operates, the Milky Way is bright and it will distort the observed map of $B$.
To reduce the contribution from the Milky Way to the measurements, we use a reference station to produce a baseline that resolves out large-scale Galactic structure \citep[e.g.,][]{Colegate2015}.
The contribution from smaller bright sources (e.g., Cassiopeia~A or Cygnus~A) cannot be completely resolved out, and is reduced by limiting the field of view (FoV) through time and frequency smearing \citep[e.g.,][]{Bridle1999}.
Moreover, the AUT and the reference station ``see'' different portions of the ionosphere, which will introduce an additional time delay between them.
The effects of the different ionosphere seen by the AUT and the reference station are calibrated using the bright point source.

\begin{table*}[ht]
\caption{\label{tab:obs} Observations used in this work.\tablefootmark{$\dagger$}}
\centering
\begin{tabular}{lcccccccc}
\hline\hline
Observation ID\tablefootmark{{\ddag}} & Antenna & \# antenna & Start time & Duration & \# TABs & TAB spacing & FoV       & Source\tablefootmark{f} \\
                                      &         & fields     &            & (s)      &         & (arcmin)    & (degrees) &        \\
\hline
L658168                  & HBA\tablefootmark{b} & 46\tablefootmark{c} & June $14$ $13$:$40$:$00$ UT  & 60  & $169$ & $1.6$ & $0.37$ & 3C147 \\
L658158                  & HBA\tablefootmark{b} & 46\tablefootmark{c} & June $14$ $13$:$30$:$00$ UT  & 60  & $169$ & $1.6$ & $0.37$ & 3C196 \\
L650445                  & LBA\tablefootmark{d} & 24\tablefootmark{e} & April $19$ $09$:$20$:$00$ UT & 600 & $271$ & $5$   & $1.32$  & 3C48  \\
L645357\tablefootmark{a} & LBA\tablefootmark{d} & 24\tablefootmark{e} & March $20$ $19$:$45$:$00$ UT & 600 & $271$ & $5$   & $1.32$  & 3C196 \\
\hline
\end{tabular}
\tablefoot{
\tablefoottext{$\dagger$}{Stations used as reference and their distance from the center of the array: RS$210$ $65$~km; RS$509$ $59$~km; RS$310$ $52$~km; and DE$605$ $226$~km.}\\
\tablefoottext{$\ddag$}{For each observation ID the three/four previous odd values contain the observations with the reference stations.}\\
\tablefoottext{a}{For this observation DE$605$ was not used as a reference.}\\
\tablefoottext{b}{Data was recorded for ten spectral windows centered at $115.0391$, $119.9219$, $124.8047$, $129.6875$, $134.5703$, $139.4531$, $144.3359$, $154.1016$, $163.8672$, and $173.6328$~MHz.}\\
 \tablefoottext{c}{All core stations except CS024.}\\
\tablefoottext{d}{Data was recorded for ten spectral windows centered at $36.7188$, $39.8438$, $45.1172$, $50.0000$, $51.5625$, $53.1250$, $56.0547$, $61.5234$, $65.2344$, and $67.7734$~MHz.}\\
 \tablefoottext{e}{All core stations.}\\
 \tablefoottext{f}{IAU names for the sources 3C147 0538+498, 3C196 0809+483, 3C48 0134+329.}
}
\end{table*}

Following the measurement equation formalism \citep{Hamaker2000}, we obtain the visibility generated by cross-correlating the signals from the AUT and the reference station as
\begin{equation}
 V_{b}^{}=J_{\mathrm{AUT},b}^{}EJ_{\mathrm{ref}}^{\dagger}\delta_{b,\mathrm{ref}},
\end{equation}
where $E$ represents the coherency matrix formed by the pure sky visibilities, $J_{\mathrm{AUT},b}$ and $J_{\mathrm{ref}}$ are respectively the Jones matrices \citep{Jones1941} of the AUT and the reference station, the subscript $b$ represents the TABs formed with the AUT, the $\dagger$ symbol denotes taking the conjugate transpose of the corresponding matrix, and $\delta_{b,\mathrm{ref}}$ is the Kroneker delta-function due to the spatial dependence of the product.
The calibration consists of finding the inverse of the visibility of the central TAB, $V_{c}^{-1}$, and right multiplying all the visibilities with it.
This is possible since $V_{c}$ is non-singular, as $E$ is non-singular by definition and the AUT measures two orthogonal polarizations.
After this, for the central beam $\tilde{V}^{}_{b=c}=V^{}_{c}V^{-1}_{c}=1$, where $1$ represents the identity matrix.
For the remaining directions $\tilde{V}^{}_{i}=V^{}_{i}V^{-1}_{c}=J^{}_{\mathrm{AUT},i}J^{-1}_{\mathrm{AUT},c}$.
This means that the calibrated visibility for the $i$-th beam only depends on the Jones matrix of the AUT, and not on the sky brightness distribution.
This relation holds if the sky coherency matrix is that of a single point-like source \citep[e.g.,][]{Smirnov2011}.
The calibrated visibilities map $B$.
The details behind the calibration method will be presented in Brentjens et al. (in prep.).

From the observed map of $B$ we determine the amplitude and phase over the aperture of the AUT, $A$.
In the far-field approximation, and for a coplanar array, they are related by \citep[e.g.,][]{D'Addario1982,Baars2007,Thompson2017},
\begin{equation}
 B(l,m)\propto\iint A(p,q)e^{2\pi i(pl+qm)\frac{\nu}{c}}dpdq,
 \label{eq:ft}
\end{equation}
where $i$ denotes the imaginary unit, $c$ is the speed of light, $\nu$ is frequency, $p$ and $q$ are orthogonal coordinates in the aperture plane, and $l$ and $m$ are the direction cosines measured with respect to $p$ and $q$.
For LOFAR, the $(p,q)$ coordinate system has its origin at the center of the aperture and it lies in the plane of the station, or in this case the plane of the \textit{Superterp} stations.
The phase of $A(p,q)$ is set, for example, by uncalibrated errors in the clock distribution, cable length, antenna position, and ionospheric phase variations across the aperture.

\section{Observations}
\label{sec:observations}

\subsection{LOFAR holography observations}
\label{ssec:lofobs}

Table~\ref{tab:obs} summarizes the observations.
Each TAB recorded complex voltages in two orthogonal polarizations (X and Y) at $5.12$~$\mu$s time resolution in ten spectral windows $195.3125$~kHz in width each.
The data were subsequently ingested into the LOFAR long-term archive.
The calibrator sources are selected to be small compared to the size of the TAB, and compared to the fringe spacing of the baselines between the AUT and the reference stations. 
The former prevents systematic distortions in the measured beam, while the latter guarantees high signal-to-noise ratios (S/N) on the baselines towards the reference stations.

The complex valued beam maps were measured on a regular hexagonal grid, $1.32$ and $0.37$ deg across for LBA and HBA, respectively.
The map size is limited by the number of TABs, spectral windows and stations that the beamformer, COBALT \citep[][]{Broekema2018}, can process simultaneously.
Per Fourier relation Equation~\ref{eq:ft} this implies a spatial resolution in the aperture plane of $270$~m (HBA at $174$~MHz) and $170$~m (LBA at $68$~MHz), comparable to the diameter of the \textit{Superterp} ($350$~m). 
The separation between TABs was set at $\lambda/D$ at the highest frequency, and kept constant for lower frequencies.
This maximizes the FoV while avoiding the overlap of aliasing artifacts with the AUT in the aperture plane, and enables simultaneous observations at different frequencies.

The required integration time is set by the error on the phase in the aperture plane, $\Delta\phi$, \citep{D'Addario1982}
\begin{equation}
 \Delta\phi\approx\dfrac{\pi D}{4\sqrt{2}d\mathrm{S/N}_{\mathrm{bm}}},
 \label{eq:dphi}
\end{equation}
where $D$ is the telescope diameter, $d$ the spatial resolution on the aperture plane, and $\mathrm{S/N}_{\mathrm{bm}}$ the signal-to-noise ratio in the complex-valued beam map; in other words, the ratio of the peak response of the array to the root mean square (rms) over the complex-valued beam map, $\mathrm{S/N}_{\mathrm{bm}}=I/\sigma$. 
Sigma can be estimated as \citep[e.g.,][]{Napier1982}
\begin{equation}
 \sigma=\sqrt{(\mbox{SEFD}_\mathrm{CS}/N_\mathrm{CS})(\mbox{SEFD}_\mathrm{RS})}/(\sqrt{\Delta\nu\Delta t}),
\end{equation}
where $N_\mathrm{CS}$ is the number of stations in the AUT; $\mbox{SEFD}_\mathrm{CS}$ and $\mbox{SEFD}_\mathrm{RS}$ are the system equivalent flux density (SEFD) of a core station and a reference station, respectively; $\Delta\nu$ is the bandwdith; and $\Delta t$ the integration time.
For the LBA the SEFD of each antenna field is $\approx30$~kJy at $60$~MHz and for the HBA $\approx3$~kJy at $150$~MHz \citep{vanHaarlem2013}.
Thus, for an observation of 3C196 with integration times of $600$~s at $60$~MHz and $60$~s at $150$~MHz we can determine the time delays with errors of $1.8$~ns and $0.4$~ns, respectively.

\begin{figure}
  \resizebox{\hsize}{!}
  {\includegraphics{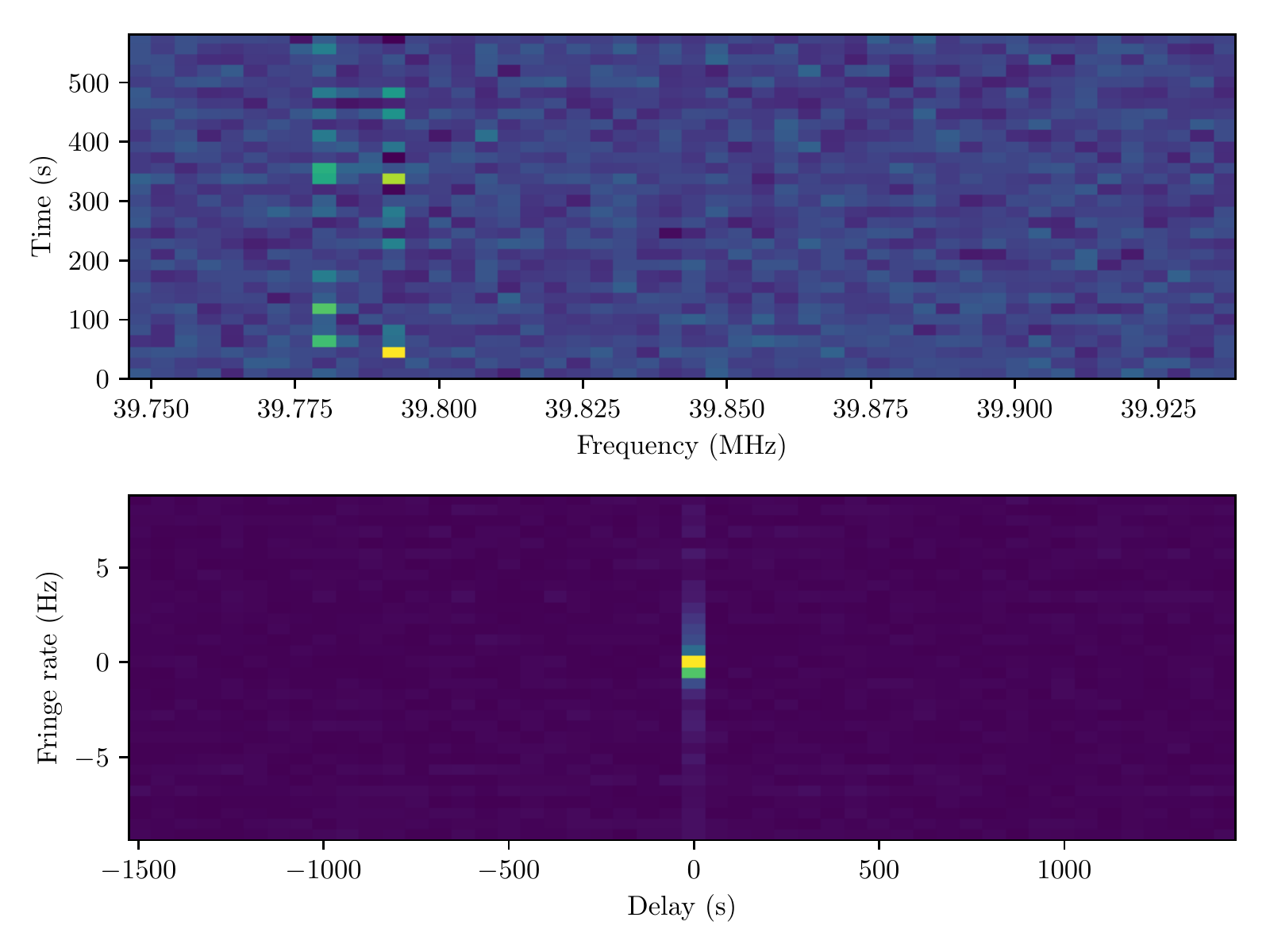}}
  \caption{Waterfall plots of the complex visibilities and their Fourier transform.
            \textit{Top panel}: Amplitude of the visibilities. The
           RFI can be seen at $39.78$~MHz.
            \textit{Bottom panel}: Fourier transform of the visibilities.
           If the visibilities are dominated by a point source in the array tracking center, then this should be a delta function centered at $0$ delay and $0$ fringe rate.
           In this case there is some time variability in the data, which produces a spread along the fringe rate.
           }
  \label{fig:waterfalls}
\end{figure}

\subsection{From raw voltages to beam maps}

To obtain a complex-valued map of the array beam we cross-correlate the voltage from the AUT with that of the reference station.
This is done using an FX correlator \citep[e.g.,][]{Thompson2017} implemented in \emph{python}.
To channelize the time series data from each spectral window we use a polyphase filter bank \citep[PFB, e.g.,][]{Price2016} with a Hann window to alleviate spectral leakage and scalloping losses.
We produce spectra of $64$ $3$~kHz channels, and a time resolution of $327.68$~$\mu$s.
This enables us to, at a later stage, flag narrowband radio frequency interference (RFI) without flagging the entire time sample.
In each spectral window we discard $25\%$ of the channels at the edges, leaving a bandwidth of $146.48$~kHz per spectral window.
The two orthogonal polarizations are combined to produce four cross-correlation products, i.e., XX, XY, YX, and YY products.

Before proceeding, we check that the bright unresolved source in the map center dominates the signal.
In this case, a time delay versus fringe rate plot will show a peaked response in the center of the diagram.
An example of such a diagram is presented in Figure~\ref{fig:waterfalls}.

\begin{figure}
  \resizebox{\hsize}{!}{\includegraphics{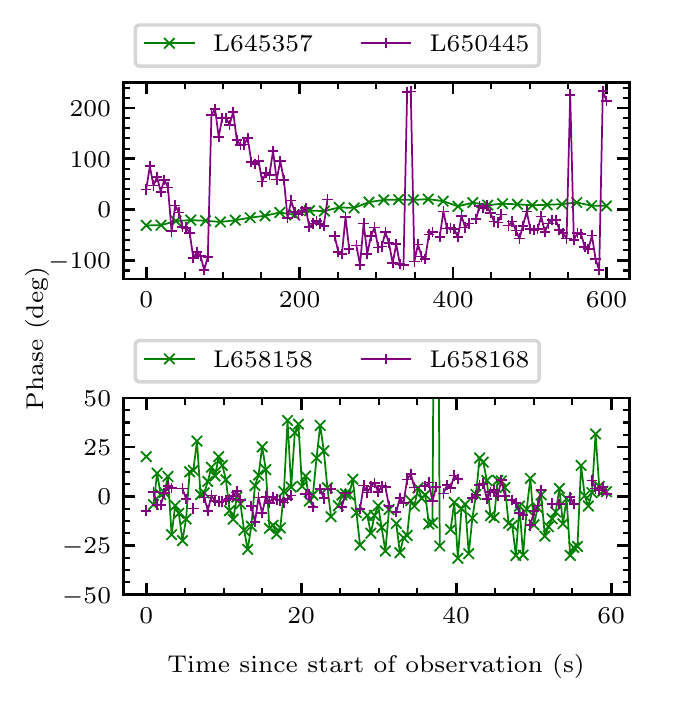}}
  \caption{Phase of the visibilities at the map center.
            \textit{Top panel}: Observed phase for the LBA observations.
            \textit{Bottom panel}: Observed phase for the HBA observations.
           \label{fig:b0phase}}
\end{figure}

After cross-correlation, the visibilities are time averaged to ensure that their S/N is high enough ($>3$) for calibration.
For the HBA observations we average to a time resolution of $0.4$~s, which results in a S/N of $6$.
For the LBA, which has a lower sensitivity and is more severely affected by the ionosphere (Figure~\ref{fig:b0phase}), the averaging times are longer.
For L645357 we average to $20$~s and for L650445 --- $5$~s.

After time averaging, we remove visibilities affected by RFI in the frequency-time domain.
We use a SumThreshold method \cite*[AOFlagger,][]{Offringa2012} on each TAB, polarization, and spectral window independently.
For the LBA and HBA observations the fraction of flagged data is $\approx5\%$.
After RFI flagging we average each spectral window in frequency to a single $146.48$~kHz channel.

The amount of time and bandwidth smearing on the visibility measured by a baseline can be approximated by \citep[e.g.,][]{Smirnov2011,Thompson2017}
\begin{equation}
 \langle V\rangle=V\sinc(\Delta\Psi)\sinc(\Delta\Phi),
\end{equation}
where $\Delta\Psi=\pi\theta_{\mathrm{s}}\Delta\nu/(\theta_{\mathrm{b}}\nu)$, $\Delta\Phi=\pi\theta_{\mathrm{s}}\omega_{\mathrm{e}}\Delta t/\theta_{\mathrm{b}}$, $\theta_{\mathrm{s}}$ is the distance from the array's phase center, $\theta_{\mathrm{b}}$ is the size of the synthesized beam formed by the baseline, and $\omega_{\mathrm{e}}$ is the Earth's rotational angular velocity ($7.2921159\times^{-5}$~radians~s$^{-1}$).
Then, for a baseline of $52$~km, a bandwidth of $146.48$~kHz, an integration time of $0.4$~s, and a frequency of $115$~MHz ($\theta_{\mathrm{b}}\approx10\arcsec$) time and bandwidth smearing reduce the amplitude of Taurus~A (the closest A-team source to 3C147) by $1.7\times10^{-4}$.
For an integration time of $5$~s, an observing frequency of $37$~MHz, and the same baseline and bandwidth the amplitude of Cassiopeia~A (closest A-team source to 3C48) is smeared by $2.3\times10^{-4}$.

The flagged and averaged visibilities are then calibrated by multiplying by the inverse of the Jones matrix of the central beam.
This has the effect of removing most of the undesired systematic effects present in the data, such as the dependence of the observed visibilities on the sky brightness distribution, beam pattern of the reference station, or ionospheric delays between reference station and AUT.

\begin{figure}[!h]
 \resizebox{\hsize}{!}
  {\includegraphics{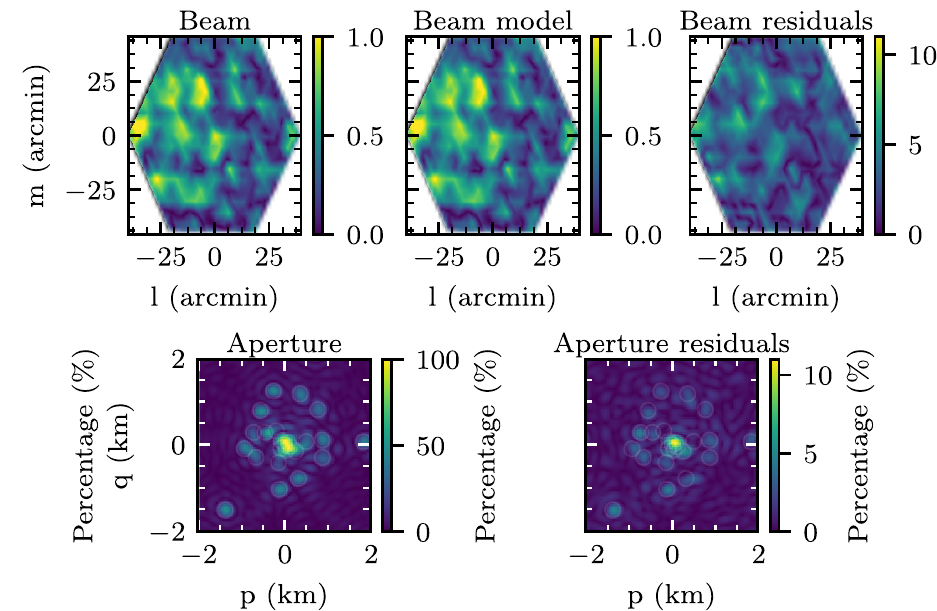}}
  \caption{Voltage beam and aperture maps for the LBA at $56$~MHz derived from the observation L645357.
           \textit{Top left}: Amplitude of the voltage beam.
           \textit{Top center}: Beam model amplitude. The beam model is generated using the derived phases for the stations in the AUT.
           \textit{Top right}: Amplitude of the residuals after subtracting the beam model from the observed voltage beam (\textit{top left}).
           \textit{Bottom left}: Aperture map obtained by taking the Fourier transform of the voltage beam (inverse Fourier transform of Equation~\ref{eq:ft}).
           The \textit{white circles} show the location of the stations in the AUT.
           \textit{Bottom right}: Amplitude of the Fourier transform of the beam residuals (\textit{top right}).
           \label{fig:bmap_lba}}
\end{figure}
\begin{figure}[!h]
 \resizebox{\hsize}{!}
  {\includegraphics{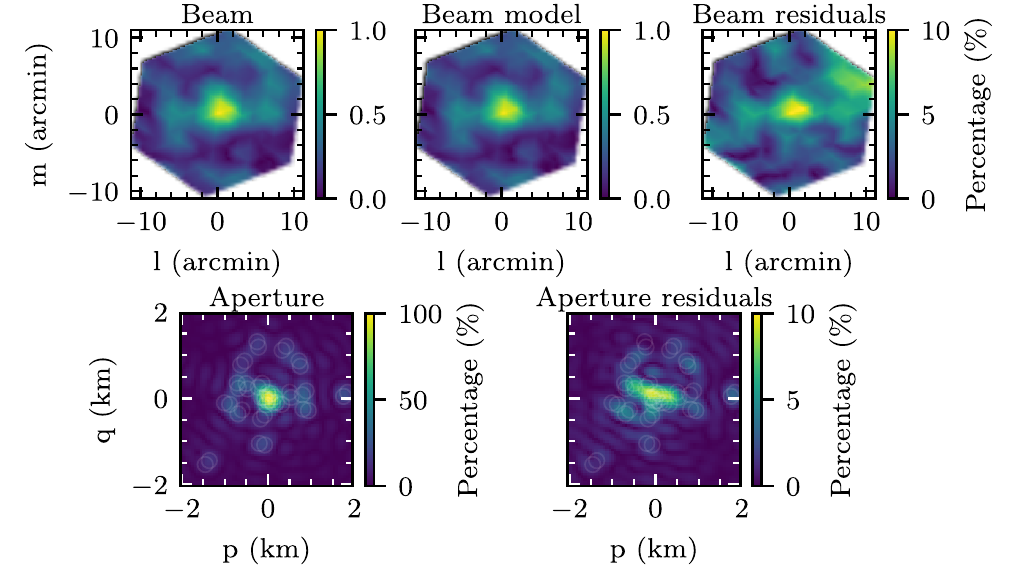}}
  \caption{Voltage beam and aperture maps for the HBA at $139$~MHz derived from the observation L658168.
           \textit{Top left}: Amplitude of the voltage beam.
           \textit{Top center}: Beam model amplitude. The beam model is generated using the derived phases for the stations in the AUT.
           \textit{Top right}: Amplitude of the residuals after subtracting the beam model from the observed voltage beam (\textit{top left}).
           \textit{Bottom left}: Aperture map obtained by taking the Fourier transform of the voltage beam (inverse Fourier transform of Equation~\ref{eq:ft}).
           The \textit{white circles} show the location of the stations in the AUT.
           \textit{Bottom right}: Amplitude of the Fourier transform of the beam residuals (\textit{top right}).
           \label{fig:bmap_hba}}
\end{figure}

After calibration, we further average the visibilities in time to one time sample with a duration of one minute for the HBA and ten minutes for the LBA (Table~\ref{tab:obs}).
After averaging in time, we are left with one calibrated complex visibility for each polarization (XX, XY, YX, and YY), spectral window, and TAB.
These calibrated complex visibilities map the complex-valued beam.
Finally, we compute the inverse-variance weighted mean beam maps, averaged over all reference stations.

\section{Results}
\label{sec:results}

\subsection{Beam and aperture maps}
\label{ssec:bamaps}

An example of the observed beam of the LBA is presented in the top left panel of Figure~\ref{fig:bmap_lba}.
There the main lobe of the beam is at the map center, and we can also see that there is a side lobe with a similar amplitude at $(l,m)=(-6,25)$.
This is produced by improperly calibrated time delays between stations.
For the HBA (top left panel of Figure~\ref{fig:bmap_hba}) the side lobes have amplitudes $\approx30\%$ of the main lobe.
The time delays between HBA stations are regularly calibrated using synthesis imaging observations.

The voltage beam is the Fourier transform of the aperture illumination (Eq.~\ref{eq:ft}), shown in the bottom left panel of Figures~\ref{fig:bmap_lba} and \ref{fig:bmap_hba}.
There we can see that the amplitudes are non-zero at the location of the stations in the AUT.
The amplitudes are larger in the \textit{Superterp} because there the stations are unresolved and their amplitudes, and phases, overlap.

\subsection{Time delays and $0$ Hz phase offsets}
\label{ssec:tau}

\begin{figure}
  \resizebox{\hsize}{!}{\includegraphics{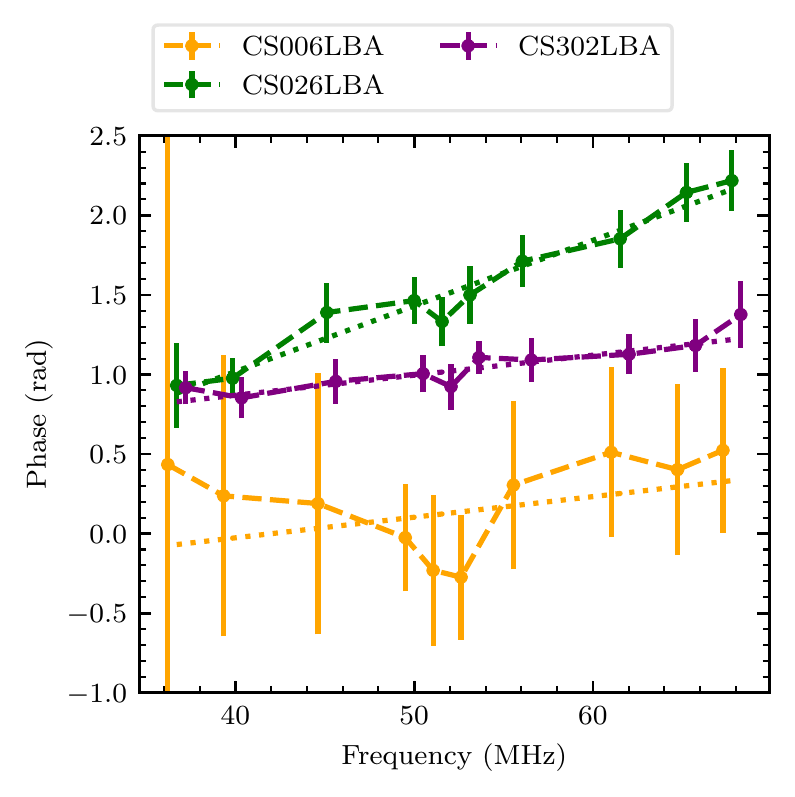}}
  \caption{Example of the measured phases as a function of frequency for three LBA stations.
           The measured phases are shown with \textit{dashed lines} and \textit{error bars}, while the best fit lines are shown with \textit{dotted lines}.
           CS302LBA is at a distance of 2 km from the array center, CS026LBA at $870$~m, and CS006LBA at $126$~m, part of the \textit{Superterp}.
           These correspond to values derived from observation L645357.
           \label{fig:phaseeg}}
\end{figure}

To measure the phase of the stations in the AUT we pose Equation~\ref{eq:ft} as a linear problem, i.e., B=Ax with
\[
 A=\begin{pmatrix}
    \exp[2\pi i(p_{s}l_{j}+q_{s}m_{j})] & \dots & \exp[2\pi i(p_{N_\mathrm{CS}}l_{j}+q_{N_\mathrm{CS}}m_{j})]\\
    \vdots & \ddots & \vdots \\
    \exp[2\pi i(p_{s}l_{N}+q_{s}m_{N})] & \dots & \exp[2\pi i(p_{N_\mathrm{CS}}l_{N}+q_{N_\mathrm{CS}}m_{N})]\\
   \end{pmatrix},
\]where $N$ is the number of TABs and x is a complex vector whose argument is the phase of each station, $\phi$.
The linear complex problem is recast to a real problem following \citet{Militaru2012}.
Then, we use least squares parameter estimation to determine the amplitude and phase at the locations of the stations.
The phases derived are not meaningful on their own, as an interferometer only measures relative phases \citep[e.g.,][]{Jennison1958}.
To remove the arbitrary offset from the phases we reference them with respect to one of the stations in the AUT.

From the referenced phases we can recover the time delay, $\tau$, and the $0$ Hz phase offset, $\phi_{0}$, of each station.
These are related to the phase by the linear relation $\phi=2\pi\nu\tau+\phi_{0}$.
An example of the observed phases and their best fit linear relation are presented in Figure~\ref{fig:phaseeg}.
There we can see that the phases show a linear relation with frequency and that the error bars on the phases become larger for stations closer to the array center.

\begin{figure}
  \resizebox{\hsize}{!}{\includegraphics{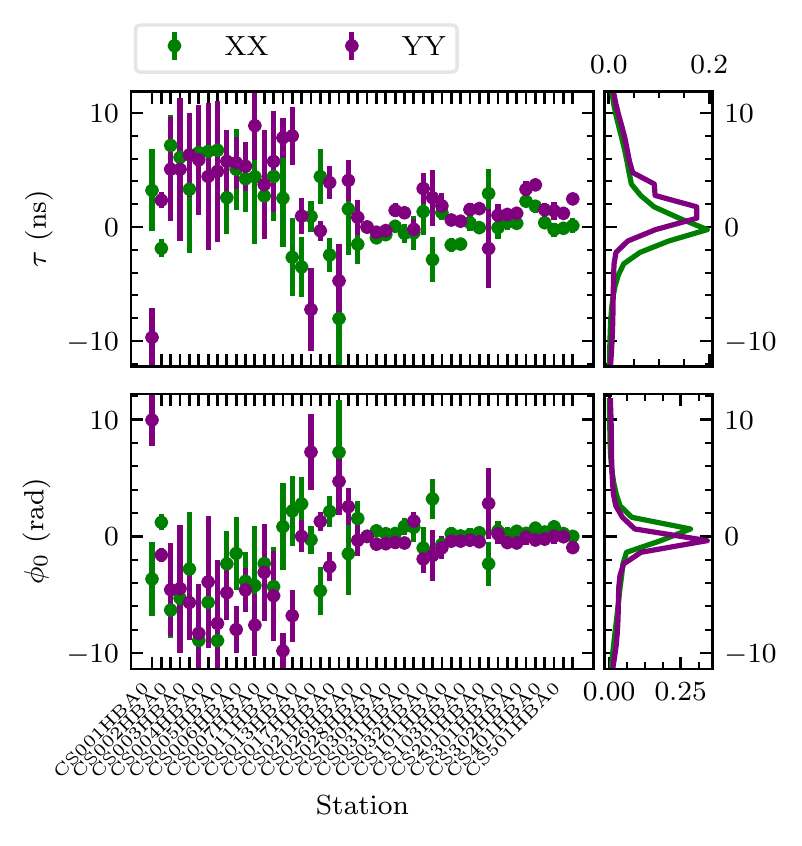}}
  \caption{Time delay and $0$ Hz phase offsets for the HBA stations derived from observation L658168.
            \textit{Top panel}: Derived time delay $\tau$ for each station in the AUT.
           The time delay for CS$026$HBA1 is $0$ because this station was used to reference the phases.
            \textit{Bottom panel}: Phase offset of $0$  Hz.
           It can be seen that the time delays and $0$ Hz phase offsets are consistent between the two polarizations.
           For the innermost stations, CS$002$ to CS$007$, the scatter is larger because the stations are unresolved.
           \label{fig:dgains_hba}}
\end{figure}

Examples of the measured $\tau$ and $\phi_{0}$ for the HBA stations derived from the L658168 observations are shown in Figure~\ref{fig:dgains_hba}.
For the \textit{Superterp} stations, CS$002$ to CS$007$, the error bars are a factor of three larger than for the rest of the stations.
This is a consequence of the larger phase errors obtained for the \textit{Superterp} stations (see Figure~\ref{fig:phaseeg}).
This is also reflected in the larger aperture residuals at the \textit{Superterp} (bottom right panel of Figure~\ref{fig:bmap_hba}).
For the stations outside the \textit{Superterp}, the errors on $\tau$ have a mean value of $1.4\pm1.2$~ns and $1.2\pm0.9$~ns for the XX and YY polarizations, respectively.
For the observation L658158 the same stations have errors on $\tau$ with a mean of $3.9\pm1.7$~ns and $3.7\pm1.7$~ns for the XX and YY polarizations, respectively.
Since the flux density of 3C196 is a factor of $1.2$ higher than that of 3C147, the larger errors on $\tau$ for L658158 are produced by the larger phase fluctuations in this observation (Figure~\ref{fig:b0phase}).

\begin{figure}
  \resizebox{\hsize}{!}{\includegraphics{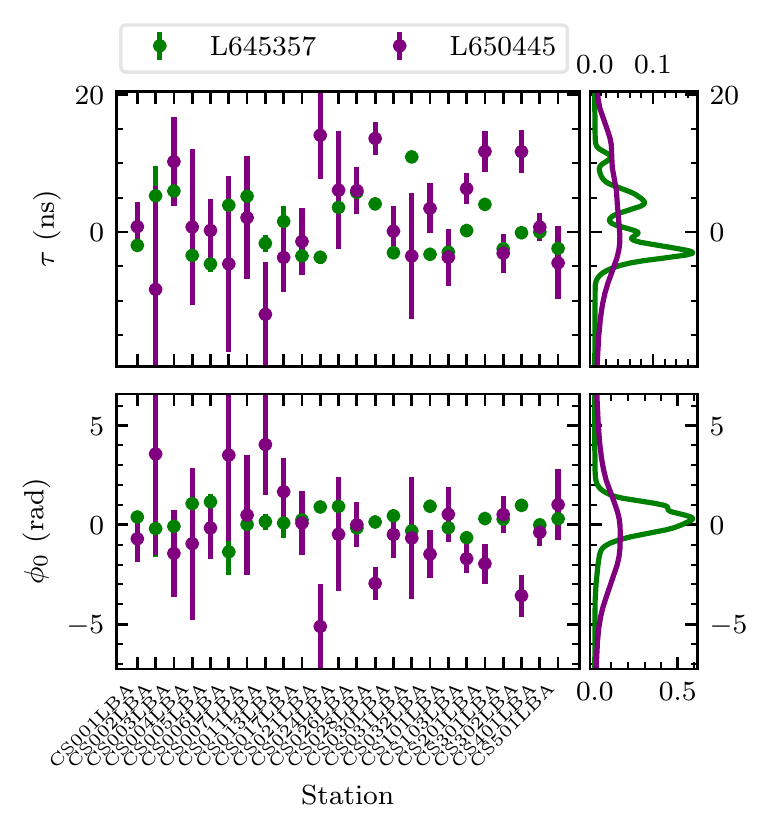}}
  \caption{Time delays and $0$ Hz phase offsets for the LBA stations derived from holography observations. 
           The time delays and $0$ Hz phase offsets correspond to those of the XX polarization for two observations (L645357 and L650445).
            \textit{Top panel}: Derived time delay $\tau$ for each station in the AUT.
           The time delay for CS$401$ is $0$ because this station was used to reference the phases.
            \textit{Bottom panel}: Phase offset of $0$ Hz.
           It can be seen that the time delays and $0$ Hz phase offsets are consistent between the two observations.
           For the innermost stations, CS$002$ to CS$007$, the scatter is larger because the stations are unresolved.
           \label{fig:dgains_lba}}
\end{figure}

The measured time delays and $0$ Hz phase offsets for the LBA stations derived from L645357 and L650445 are presented in Figure~\ref{fig:dgains_lba}.
In both observations the derived values of $\tau$ and $\phi_{0}$ agree to within $3\sigma$, even though in L645357 the S/N is higher by a factor of $12$; for the L645357 observations the ionosphere over the array produces a smooth slow time-varying phase rotation, while for L650445 the changes are faster and more pronounced (Figure~\ref{fig:b0phase}).
The time delays for the \textit{Superterp} stations have errors that are a factor of four larger than for the rest of the stations.

\begin{figure}
  \resizebox{\hsize}{!}{\includegraphics{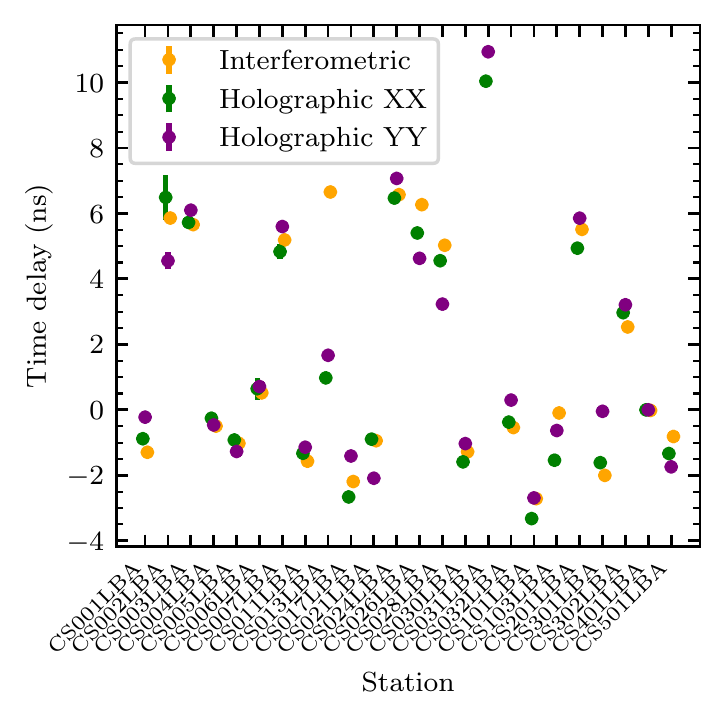}}
  \caption{Time delays for the LBA stations assuming that $\phi_{0}=0$.
           The time delays obtained from holography observations are shown for two orthogonal polarizations (XX and YY).
           We also show the time delays derived from imaging observations for Stokes I.
           During the calibration of imaging observations the phase offset between XX and YY polarizations is removed by assuming that a reference station has zero phase offsets \citep[][]{deGasperin2019}.
           The time delay for CS$401$LBA is $0$ because this station was used to reference the phases.
           Station CS$031$LBA is missing in the imaging data because it was completely flagged.
           \label{fig:tau_lba_fix}}
\end{figure}

For both HBA and LBA (Figures~\ref{fig:dgains_hba} and \ref{fig:dgains_lba}), the $0$ Hz phase offsets are consistent with being zero at the $5\sigma$ level.
Motivated by this, we fit a linear relation to the phases with $\phi_{0}=0$.
The values of $\tau$ for the LBA stations under this assumption are presented in Figure~\ref{fig:tau_lba_fix}.
We can see that the derived time delays are consistent with those presented in Figure~\ref{fig:dgains_lba}, but in this case the error bars are smaller because there is one less free parameter and setting $\phi_{0}=0$ is a strong constraint.
Using $\phi_{0}=0$ the mean value of the error of the derived time delays is $0.26\pm0.16$~ns and $0.17\pm0.10$~ns for the HBA and LBA, respectively.

In Figure~\ref{fig:tau_lba_fix} we also show the time delays for the LBA stations derived from imaging observations.
In imaging observations the phases for each station are derived from observations of a bright calibrator source and a model of the sky brightness distribution \citep[e.g.,][]{Fomalont1999}.
Then the contribution to the phase from the station delays and the ionosphere are separated \citep[e.g.,][]{vanWeeren2016,deGasperin2018,deGasperin2019}.
We see that the time delays derived using holography and imaging observations agree to within $3\sigma$ for $20$ out of the $23$ stations present in both observations.
This shows that the time delays derived here, where no model of the sky brightness distribution is used, are indistinguishable from those derived in imaging observations.
The interferometric time delays have smaller error bars because they are obtained using $488$ $195.3125$~kHz spectral windows.

We check that the derived time delays and $0$ Hz phase offsets capture the status of the AUT by using them to simulate the array beam and comparing it with the observed beam.
To simulate the array beam we use the time delays and $0$ Hz phase offsets shown in Figures~\ref{fig:dgains_hba} and \ref{fig:dgains_lba} for the HBA and LBA stations, respectively.
These are used to evaluate the phase of each station at a particular frequency, producing a complex valued map of the AUT.
The Fourier transform of the AUT simulates the array beam.
The residuals between the observed and simulated array beams are presented in the top right panel of Figures~\ref{fig:bmap_lba} and \ref{fig:bmap_hba}.
The residuals in the image plane have no obvious structure and show amplitudes of $\lesssim10\%$.
This shows that we can reproduce the array beam using the derived time delays and $0$ Hz phase offsets.
However, the Fourier transform of the beam residuals reveals that there is significant structure in the aperture plane (bottom right panel of Figures~\ref{fig:bmap_lba} and \ref{fig:bmap_hba}).
This can be seen as a larger amplitude ($11\%$ for the LBA) at the location of the \textit{Superterp}, for which we are not capturing the phase behavior as accurately as for the stations away from it (see also Figure~\ref{fig:phaseeg}).

\section{Discussion}
\label{sec:discussion}

\subsection{Corrected time delays}
\label{ssec:cortau}

\begin{figure}
  \resizebox{\hsize}{!}{\includegraphics{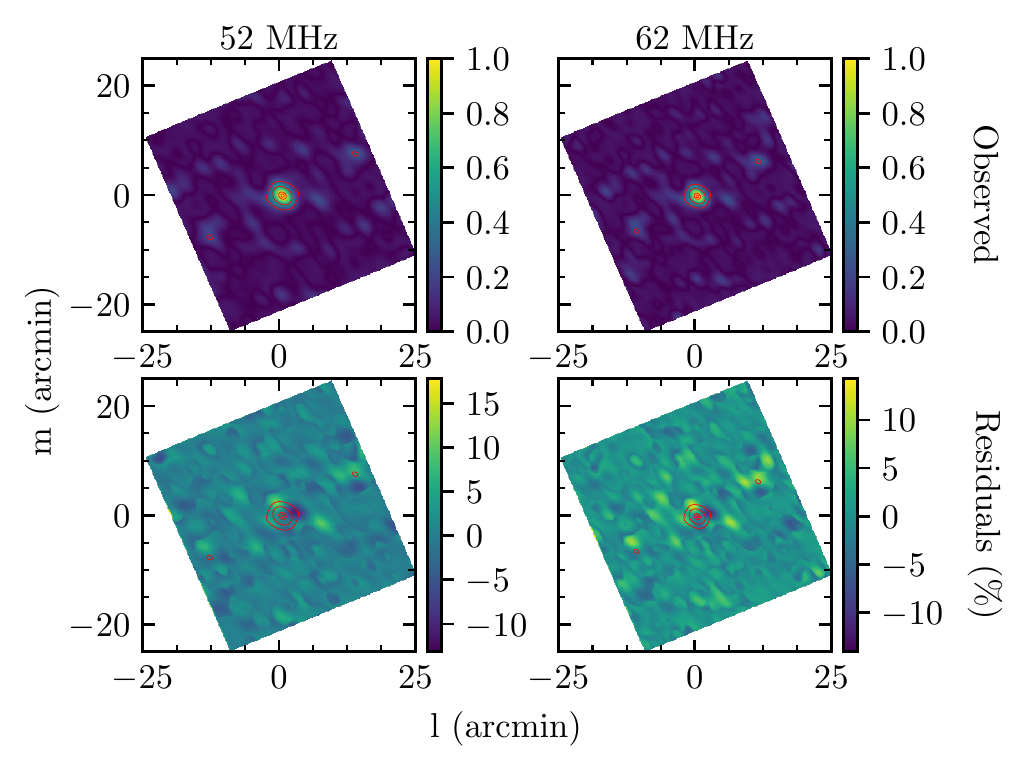}}
  \caption{Observed beam for the LBA.
           The \textit{top row} shows the observed beam after updating the instrumental time delays in the beamformer (\textit{colorscale}).
           The \textit{red contours} show the amplitude of a beam model generated from the position of the stations with no additional time delays.
           They are at the $20\%$, $50\%$, $90\%$ and $98\%$ level with respect to the peak response.
           The \textit{bottom row} shows the residuals after subtracting the beam model from the observed beam (\textit{colorscale}).
           The \textit{red contours} are the same as in the top row.
           \label{fig:corrbeam}}
\end{figure}

We use the derived time delays (Figure~\ref{fig:tau_lba_fix}) to update the instrumental time delays in LOFAR's beamformer. 
To test the effect of updating the instrumental time delays in the beamformer we observed Cygnus~A with the core stations of LOFAR.
The observation was one minute long using the imaging mode, where the signals of different stations are cross-correlated instead of added.
Since the signal path between stations and the beamformer is the same in tied-array and imaging modes, these observations have a beam equivalent to the one observed using holography.
Cygnus~A has a size of $\approx1\arcmin$ \citep[e.g.,][]{McKean2016}, so it will be unresolved by the LOFAR core at LBA frequencies (at $90$~MHz the spatial resolution of the core is $3\farcm3$).
Hence, a dirty image obtained from this observation will show the array beam.

A comparison between the LBA beam after the update and a model of the beam is presented in Figure~\ref{fig:corrbeam}.
In the beam model the phase of each station is given by its location with no additional time delays.
We see that the sidelobes in the observed beam are similar to those of the beam model.
After subtracting the beam model the residuals are $\lesssim20\%$.

\begin{figure}
  \resizebox{\hsize}{!}{\includegraphics{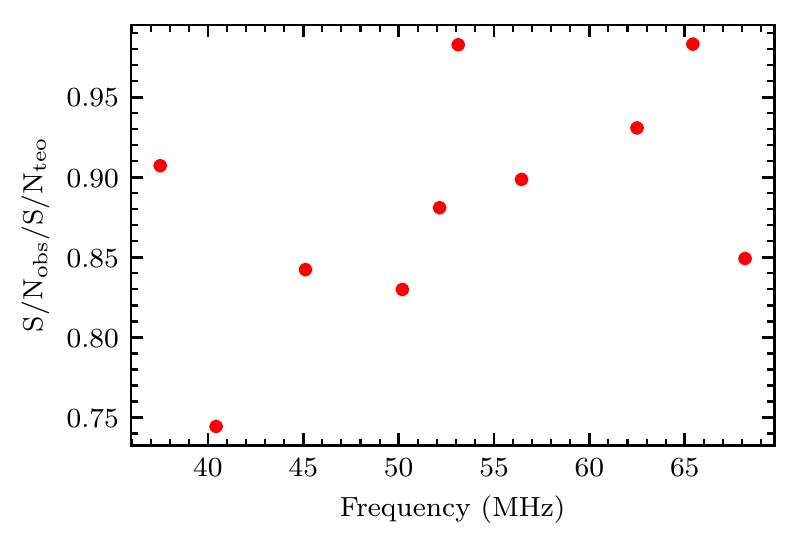}}
  \caption{Ratio of the observed S/N to the theoretical value.
           \label{fig:snr_ratio}}
\end{figure}

We compare the S/N of the observations of Cygnus~A with the theoretical S/N in Figure~\ref{fig:snr_ratio}.
The ratio of the observed S/N to its theoretical value has a mean of $0.88\pm0.06$.
These results show that after updating the time delays in the beamformer the array beam is close to the ideal case.
The remaining differences are produced by propagation delays introduced by the ionosphere and any remaining errors in the instrumental time delays.

When using an unpolarized bright unresolved source to determine the time delays, there will be an arbitrary offset between the X and Y polarizations.
When these time delays are applied to the beamformer, the offset will produce a rotation of Stokes $U$ into Stokes $V$.
In order to find this offset we need to observe a linearly polarized source for which the sign of the rotation measure is known.
From this observation the offset is determined from the angle between the apparent Stokes $U$ and $V$ \citep[e.g.,][]{Brentjens2008}.
This step has not been performed yet.

\subsection{Comparison to other methods}

Previous to holography LOFAR used interferometric observations to derive time delays between its stations using the methods described by \citet{Wijnholds2009}.
In these interferometric array-calibration observations pairs of antennas were cross-correlated using the station correlator.
The antenna gains were derived by calibrating against multiple calibrator sources in their FoV.
The observations lasted $6$~hours and $24$~hours for the HBA and LBA, respectively.
Variations in the sky brightness distribution due to the ionosphere were partially averaged out during the observations \citep[e.g.,][]{Wijnholds2011}.
By comparison, holography requires $1$~minute and $10$~minutes of observations for the HBA and LBA, respectively, and it does not require the use of a sky model.

Time delays between stations are also derived during imaging observations.
In the case of Figure~\ref{fig:tau_lba_fix} the time delays derived from imaging observations have smaller error bars because they are derived from $488$ $195.3125$~kHz spectral windows.
If we derive time delays from the imaging observations using the same ten spectral windows as for the holographic observations, then the errors on the time delays have a difference of less than $5\%$.
This makes holography a competitive alternative, as its accuracy can be scaled up by adding more reference stations and more spectral windows.
For the latter an increase in computing power is required.

The method presented here is also used to calibrate the antennas within a LOFAR station.
For this calibration a station is under test (instead of the AUT).
The station under test generates multiple station-beams to map its beam, while another station acts as reference.
The complex voltages from the station under test and reference are then cross-correlated and calibrated following the same procedure as that outlined here.
From the calibrated complex visibilities the complex gains for each antenna within a station are derived.

\subsection{Improvements to holographic measurements with LOFAR}

One of the main limitations of the holographic measurements presented is the spatial resolution over the telescope aperture.
For a constant number of TABs this can be improved by observing a larger portion of the beam using a mosaic while keeping a TAB at the calibrator source.
Additionally, the separation between TABs could be made smaller, reducing aliasing artifacts in the aperture plane.

For the experiments presented in this work we used only four reference stations.
Outside its core, LOFAR has $14$ stations within the Netherlands and $13$ stations distributed all over Europe.
Any of these stations can be used as reference station, as long as the baseline formed with the AUT does not resolve the source used to map the beam.
This means that there can be an improvement in the S/N of the complex-valued beam map of up to a factor of four using the same sources.
With this level of improvement in the maps of the complex-valued beam, the integration times could be made shorter or more precise time delays could be derived.

The time delays derived from holographic measurements can be used to update the instrumental time delays prior to an observation with the tied array (e.g., of a pulsar).
Moreover, if the integration time required to reach nanosecond precision could be made shorter, and the post-processing of the holographic measurements could be done in real time, then it would be possible to interleave holographic observations during the tied-array observations.
This could be a way towards semi-real-time beam calibration for the Square Kilometer Array (SKA).
Implementing a dedicated holography mode in the supercomputer that processes the raw LOFAR data is one of the next steps towards this goal.

\section{Summary}
\label{sec:summary}

In this work we used radio holography along with a new calibration method to characterize the time delays between LOFAR's core stations.
This new calibration method consists in calibrating the measured complex-valued beam map by right multiplying by the matrix inverse of the map center.
This calibration makes the observed complex-valued reception pattern independent of the sky brightness distribution.

Four HBA and three LBA reference stations were used simultaneously to produce maps of the tied-array voltage beam.
Using $60$~s (HBA) and $600$~s (LBA) long observations of 3C196, 3C147, and 3C48 we derived time delays with an error $<1$~ns.
We find that the main limitations in reaching nanosecond precision in the measured time delays are the condition of the ionosphere over the array and the ability to spatially resolve the array elements.
LOFAR now uses the derived time delays operationally.

\begin{acknowledgements}
      P.~S., J.~B.~R.~O. and H.~J.~A.~R. acknowledge financial support from the Dutch Science Organisation (NWO) through TOP grant 614.001.351.
      This research made use of Astropy, a community-developed core Python package for Astronomy \citep{Astropy2013}, and matplotlib, a Python library for publication quality graphics \citep{Hunter2007}.
\end{acknowledgements}

\bibliographystyle{aa}
\bibliography{lofar_holog.bib}

\end{document}